\documentclass[aps,pra,twocolumn,showpacs]{revtex4}
\usepackage{amsmath, amssymb}
\usepackage{color, graphicx}

\newtheorem{corollary}{Corollary}
\newtheorem{definition}{Definition}
\newtheorem{theorem}{Theorem}

\begin{document}
\title{Determination of the Schmidt number}

\author{J. Sperling} \email{jan.sperling2@uni-rostock.de}
\author{W. Vogel} \email{werner.vogel@uni-rostock.de}
\affiliation{Arbeitsgruppe Quantenoptik, Institut f\"ur Physik, Universit\"at Rostock, D-18051 Rostock, Germany}

\date{\today}

\begin{abstract}
Optimized, necessary and sufficient conditions for the identification of the Schmidt number will be derived in terms of general Hermitian operators.
These conditions apply to arbitrary mixed quantum states.
The optimization procedure delivers equations similar to the eigenvalue problem of an operator.
The properties of the solution of these equations will be studied.
We solve these equations for classes of operators.
The solutions will be applied to phase randomized two-mode squeezed-vacuum states in continuous variable systems.
\end{abstract}
\maketitle

\section{Introduction}
Entanglement is the key resource of the vast fields of Quantum Information Processing, Quantum Computation, and Quantum Technology, for an introduction see e.g.~\cite{book1,book2}.
For example, applications of entangled states are those for quantum key distribution~\cite{ekert91}, quantum dense coding~\cite{bennett-wiesner92}, and quantum teleportation~\cite{bennett93}.
Thus both, the identification and the quantification of entanglement, play a mayor role for future applications~\cite{book3}.

The phenomenon entanglement is closely related to the superposition principle of quantum mechanics.
A pure separable state is represented by a product of states for both systems.
A general pure state is a superposition of factorizable states.
The minimal number of such global superpositions denotes the Schmidt rank~\cite{book1}.
A separable mixed quantum state is a convex combination of pure factorizable quantum states~\cite{Werner}.
The generalization of the Schmidt rank to mixed quantum states delivers the Schmidt number (SN).
This generalization and the introduction of SN witnesses is given in~\cite{Sanpera2,Terhal,Bruss}.
The experimental construction of states with a certain SN has been realized in~\cite{URen}.
The SN of a mixed quantum state fulfills the axioms of an entanglement measure, cf.~\cite{Vedral,Vidal2,Vedral2}.
More precisely, it is a convex roof measure as defined in~\cite{Uhlmann,Bennett}.

The identification of entanglement in terms of entanglement witnesses has been introduced in~\cite{physlettA223-1}.
For a given witness an optimization can be performed~\cite{physrevA62-052310}.
Recently, we proposed optimized, necessary and sufficient conditions for the detection of entanglement~\cite{SpeVo1}.
Note that the latter optimization and the optimization of an entanglement witness are inherently different.
Our optimization procedure delivers so-called separability eigenvalue equations.
They resemble the well-known eigenvalue problem, but they include the factorization property of pure local quantum states.

In the present contribution we study the identification of the SN of a given quantum state.
We provide a method that deliver all optimized SN witnesses.
The optimization yields equations which will be discussed in two forms.
These generalizations of the separability eigenvalue problem delivers similar equations for arbitrary mixed SN states.
Our method will be compared with the well known spectral decomposition and the Schmidt decomposition of quantum states.
Properties of the solutions of these equations will be studied.
With some fundamental examples, we generate some general classes of SN witnesses.
We apply them to mixed quantum states.

The paper is structured as follows.
We motivate our method in Sec.~\ref{Sec:Motiv}.
In Sec.~\ref{Sec:Witness} we reformulate the detection of the SN of a quantum state by witnesses in terms of arbitrary Hermitian operators and optimized, necessary and sufficient conditions.
The optimization procedure will be discussed in Sec.~\ref{Sec:SNEVAL}, where we express the optimization problem in terms of a perturbed eigenvalue problem and discuss properties of these equations.
In Sec.~\ref{Sec:SESol} we solve these equations for a wide class of operators, including all one dimensional projectors and operators defined in continuous variables.
We apply the method in Sec.~\ref{Sec:Ex} to identify a SN greater than one and two, for the case of a phase diffused two-mode squeezed-vacuum state.
A summary and some conclusions are given in Sec.~\ref{Sec:SC}.

\section{Motivation}\label{Sec:Motiv}
\begin{center}
\begin{figure}[ht]
\includegraphics*[width=6cm]{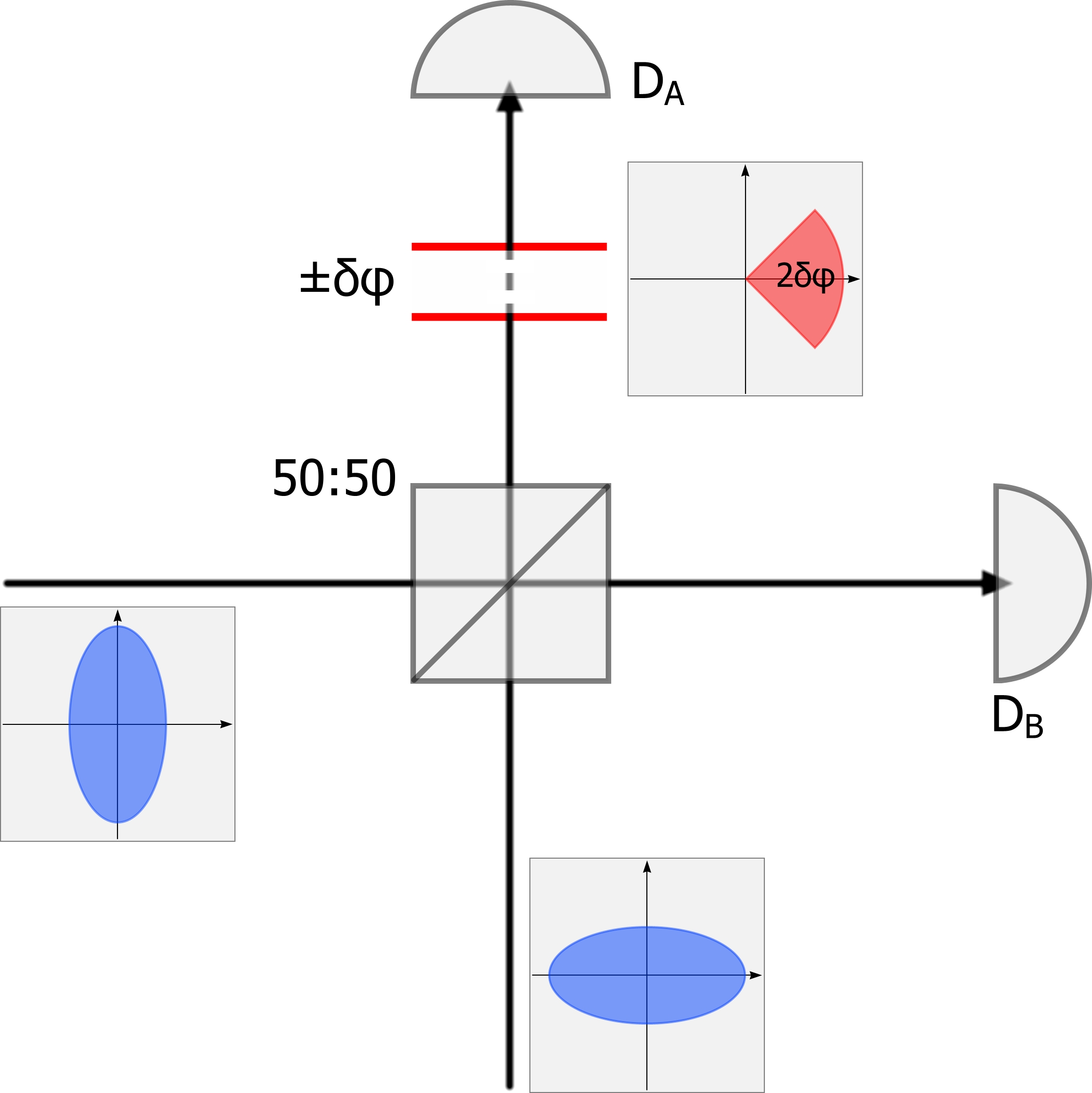}
\caption{
(color online) The experimental realization for the considered setup is the following.
In both beam splitter inputs, we have squeezed-vacuum states.
The squeezing is considered in orthogonal quadratures.
The 50:50 beam splitter delivers an entangled two-mode squeezed-vacuum state.
The detectors ${\rm D}_A$, ${\rm D}_B$ resemble the measurement of a, so far unknown, test operator $L$.
The fluctuation of the distance between the beam splitter output and the detector ${\rm D}_A$ yield the phase randomization $\pm\delta\varphi$.
The distance between the beam splitter output and the detector ${\rm D}_B$ is fixed.
}\label{Fig:Exp}
\end{figure}
\end{center}

Let us consider the following experimental situation, cf. Fig.~\ref{Fig:Exp}.
We have a beam splitter with squeezed-vacuum states in both inputs.
These states have the same amount of squeezing, but in orthogonal quadratures.
The output is the two-mode squeezed-vacuum state $|q\rangle$, ($q=\epsilon e^{i\varphi}$ and $0<\epsilon<1$),
\begin{align}\label{Eq:1}
	|q\rangle=\sqrt{1-|q|^2}\sum_{k=0}^\infty q^k |k,k\rangle.
\end{align}
One output is disturbed in terms of a phase randomization, which equally randomizes the phase $\varphi$ between $-\delta\varphi$ and $+\delta\varphi$.
The scenario under consideration could be used for transferring one part of an entangled state through a noisy channel.
The sender keeps the other part of the state, e.g. in a delay line, such as an optical fiber.

The measured state is given by
\begin{align}
	\nonumber \rho_{\delta\varphi}=&\frac{1}{2\delta\varphi}\int_{-\delta\varphi}^{+\delta\varphi} d\varphi\,\left(U(\varphi)\otimes\mathbb I\right)|\epsilon\rangle\langle\epsilon|\left(U(\varphi)\otimes\mathbb I\right)^\dagger\\
	=&\frac{1}{2\delta\varphi}\int_{-\delta\varphi}^{+\delta\varphi} d\varphi\, |\epsilon e^{i\varphi}\rangle\langle\epsilon e^{i\varphi}|,\label{Eq:RhoVarPhi}
\end{align}
with the local unitary operation $U(\varphi)=\sum_{k=0}^\infty e^{i\varphi k}|k\rangle\langle k|$.
In general, this state is a mixed quantum state in continuous variable systems.
Due to the linear independence of the two-mode squeezed-vacuum states, the rank of the operator $\rho_{\delta\varrho}$ is, in general, also infinite.

The two-mode squeezed-vacuum state has been considered, for example, as a recource for quantum teleportation~\cite{TMSVuse1,TMSVuse2}, quantum dense coding~\cite{TMSVuse3}, and quantum memories~\cite{TMSVuse4}.
It has been shown that this state violates a continuous variable Bell inequality~\cite{TMSVuse5}.
Pertubations of the pure two-mode squeezed-vacuum state have been studied, such as phase and amplitude damping~\cite{TMSVuse6}, or noise due to the transmission in optical fibers~\cite{TMSVuse7}.
Here, we focus on the SN of the phase randomized two-mode squeezed-vacuum state $\rho_{\delta\varphi}$.

In the case of zero phase randomization, we have an infinite SN for $\rho_0=|\epsilon\rangle\langle\epsilon|$.
This state includes an entangled qubit ($k=0,1$), an entangled qutrit ($k=0,1,2$), \dots, an entangled qudit ($k=0,1,\dots,d-1$), see Eq~(\ref{Eq:1}).
The two-mode squeezed-vacuum state is a global superposition of infinitely many product states $|k,k\rangle$, which corresponds an infinite SN.
In this case, entanglement determines all the correlation between system $A$ and system~$B$.

Now let us consider the fully randomized state $\delta\varphi=\pi$,
\begin{align}
	\nonumber \rho_\pi&=\frac{1-\epsilon^2}{2\pi}\sum_{k,l=0}^\infty\epsilon^k\epsilon^l\int_{-\pi}^{+\pi} d\varphi\, e^{ik\varphi}e^{-il\varphi}|k,k\rangle\langle l,l|\\
	&=(1-\epsilon^2)\sum_{k=0}^\infty \epsilon^{2k} |k,k\rangle\langle k,k|.\label{Eq:TMSVstate}
\end{align}
This state is obviously separable.
Thus, it does not include any entangled qudits, and the SN is one.
The subsystems $A$ and $B$ are only classically correlated.

Some questions arise automatically.
Somewhere in between the two extreme cases, $\delta\varphi=0$ and $\delta\varphi=\pi$, all the qudits disappear one after another, and the quantum correlations between $A$ and $B$ vanish.
We aim to answer questions such as: {\em For which value $\delta\varphi$ the state becomes separable?}, or: {\em Which phase randomization $\delta\varphi$ still deliver a quantum state with, for example, an entangled qutrit?}
With other words, we want to quantify the entanglement of the state $\rho_{\delta\varphi}$ under a certain phase randomization by the SN.
Let us divide the problem into several sub-problems, which we solve in this manuscript:

\paragraph{Conditions.}
We will derive general, necessary and sufficient conditions for the identification of a certain Schmidt number $r$ (SN-$r$).
This method applies to all mixed quantum states.
These conditions are based on measuring a general observable $L$, $\langle L\rangle$.
For obtaining the SN of the state, we relate $\langle L\rangle$ to the maximal expectation value under all quantum states with a SN less or equal to $r$, $f_r(L)$.

\paragraph{Optimization.}
We derive equations in two different forms, to obtain the desired function $f_r$.
These equations resemble the eigenvalue problem.
The global superposition property of SN-$r$ quantum states is encoded in these equations.

\paragraph{Solution.}
We analyze these new kinds of equations.
We find a close relation to the Schmidt decomposition and the spectral decomposition of an operator.
Afterward, we solve them for a large class of operators - including those for our considered example.

\paragraph{Applying the method.}
We apply our method to the phase randomized two-mode squeezed-vacuum state.
We show how the squeezing and the phase randomization influence the quantum correlations between subsystems $A$ and $B$.
We identify entanglement in general and the existence of a qutrit in the randomized state.

Altogether, we derive a new mathematical method for the identification of the SN.
This method allows us to test up to which order entangled qudits are contained in an arbitrary bipartite quantum state.
These entangled qudits can be used for quantum information processing.

\section{Schmidt number states and witnesses}\label{Sec:Witness}
Let us consider a bipartite quantum system which is given by compound Hilbert space $\mathcal H=\mathcal H_A\otimes \mathcal H_B$.
Previously we have shown in, how entanglement in cantinuous variable systems can be identified in finite dimensional spaces~\cite{SpeVo3}.
Thus, it is not a restriction if we assume $\dim\,\mathcal H=d_Ad_B<\infty$.
In addition let us denote the sets of linear operators acting on the Hilbert space as ${\sf Lin}(\mathcal H,\mathcal H)$, and the Hermitian operators as ${\sf Herm}(\mathcal H)$.

Statistical mixtures of pure factorizable states define the separable quantum states~\cite{Werner}.
Following this idea, quantum states with a SN less or equal to $r$ are mixtures of pure states with a Schmidt rank less or equal to $r$.
Pure states $|\psi\rangle$ with a SN less or equal to $r$, $r(\psi)\leq r$, are elements of the set $\mathcal S_r^{\rm (pure)}$ and can be written as~\cite{book1}
\begin{align}\label{Eq:SchmDec}
	|\psi\rangle=U\otimes V\sum_{k=1}^{r(\psi)} \lambda_k |k,k\rangle=\sum_{k=1}^{r(\psi)} \lambda_k |e_k,f_k\rangle,
\end{align}
with a local unitary operation $U\otimes V$, orthonormal $|k\rangle$, $|e_k\rangle$, and $|f_k\rangle$, and Schmidt coefficients $\lambda_k>0$.
The mixed SN-$r$ states $\sigma$ are elements of $\mathcal S_r$ given as convex combinations of those pure ones, $\mathcal S_r={\sf conv}\,\mathcal S_r^{\rm (pure)}$.
Let us note that the maximal possible SN is $d = {\rm min}\{ d_A ,d_B\}$, and the minimal one is $r=1$ for separable quantum states.

Analogously to entanglement witnesses, a SN witnesses $W$ is a Hermitian operator, with
\begin{align}
	&\forall \sigma\in\mathcal S_r: {\rm Tr}\,\sigma W\geq0,\label{Eq:SNWit}\\
	&\exists \varrho\in\mathcal S_{d}: {\rm Tr}\,\varrho W<0.\label{Eq:WitWit}
\end{align}
An operator fulfilling Eq.~(\ref{Eq:SNWit}) is called SN-$r$ witness, $W\in\mathcal W_r$.
The condition given in Eq.~(\ref{Eq:WitWit}) will be considered at the end of our treatment.
Here, let us only note that the set $\mathcal W_r$ includes all SN-$r$ witnesses fulfilling Eq.~(\ref{Eq:WitWit}), and all positive semi-definite operators.
Here, a SN witness $W$ is considered to be optimal, if there exits a SN-$r$ state $\sigma\in\mathcal S_r$ with ${\rm Tr}\,\sigma W=0$.

\subsection{Conditions for Schmidt number states}
In this subsection we aim to derive the conditions for the identification of SN-$r$ states.
It has been considered in~\cite{Sanpera2,Terhal,Bruss}, that a quantum state has a SN greater than $r$, if and only if
\begin{align}
	\exists W\in\mathcal W_r: {\rm Tr}\,\varrho W<0.
\end{align}
The existence of such a witness is a consequence of the Hahn-Banach Theorem.
Whereas, the general structure of SN-$r$ witnesses -- elements of $\mathcal W_r$ -- is not clear.
Thus, it is an advantage to find conditions in terms of arbitrary Hermitian operators with a well-known structure.

For this purpose let us define the following function $f_r$.
This function maps a Hermitian operator to its maximal expectation value of all mixed and pure SN-$r$ states.
\begin{definition}\label{Def:MaxExp}
The maximal SN-$r$ expectation value is given by
\begin{align*}
	f_r(L)=\sup \{\langle \psi|L|\psi\rangle: |\psi\rangle\langle\psi|\in\mathcal S_r^{\rm (pure)}\}.
\end{align*}
\end{definition}
\paragraph*{Proof.}
This value exists due to the compactness of $\mathcal S_r^{\rm (pure)}$.
This value is the maximal expectation value of {\em all} SN-$r$ states:
\begin{itemize}
	\item[''$\leq$'':] $\mathcal S_r^{\rm (pure)}\subset\mathcal S_r$ $\Rightarrow$ $f_r(L)\leq\sup \{{\rm Tr}\,\sigma L: \sigma\in\mathcal S_r\}$
	\item[''$\geq$'':] $\sigma=\sum_{|\psi\rangle\langle\psi|\in\mathcal S_r^{\rm (pure)}} p_\psi |\psi\rangle\langle\psi| \in\mathcal S_r$, with $p_\psi\geq0$ and $\sum_\psi p_\psi=1$
	\begin{align*}
		f_r(L)=&\sum_{\psi} p_\psi f_r(L)\\
		\geq & \sum_{|\psi\rangle\langle\psi|\in\mathcal S_r^{\rm (pure)}} p_\psi \langle \psi|L|\psi\rangle={\rm Tr}\,L\sigma
	\end{align*}
	$\Rightarrow$
	$f_r(L)\geq\sup \{{\rm Tr}\,\sigma L: \sigma\in\mathcal S_r\}$
\end{itemize}
It follows $f_r(L)=\sup \{{\rm Tr}\,\sigma L: \sigma\in\mathcal S_r\}$.
$\blacksquare$

In the following, the decomposition of all operators of the set $\mathcal W_r$ in terms of arbitrary Hermitian operators is considered.
We show that a certain form delivers a SN-$r$ witness, and all SN-$r$ witness can be written in this form.
This yields the identification of the SN.

Let us consider an arbitrary Hermitian operator $L\in{\sf Herm}(\mathcal H)$, and a number $\lambda\geq f_r(L)$.
We define the operator $W_L$ as
\begin{align}\label{Eq:SNWitDec}
	W_L=\lambda\mathbb I\otimes\mathbb I-L,
\end{align}
with the bipartite identity operator $\mathbb I\otimes\mathbb I$.
This is a generalization of the construction of entanglement witnesses, cf.~\cite{SpeVo1,Toth2005}.
Obviously holds for all SN-$r$ quantum states $\sigma$
\begin{align}\label{Eq:SNPos}
	{\rm Tr}\,\sigma W_L=\lambda {\rm Tr}\,\sigma-{\rm Tr}\, \sigma L\geq \lambda-f_r(L)\geq0.
\end{align}
Thus, $W_L$ is a SN-$r$ witness.

In the case $\lambda=f_r(L)$, we call this SN-$r$ witness optimized.
For $r=1$, our optimality condition is, in general, weaker then the definition in Ref.~\cite{physrevA62-052310}.
To be precise, our optimized witness, $\lambda=f_r(L)$, is finer than the witness for $\lambda>f_r(L)$.

The other way around we decompose all witnesses in this form.
Let $W\in\mathcal W_r$ be a SN-$r$ witness.
Using Eq.~(\ref{Eq:SNWit}) multiplied with $-1$ and the resulting choice $\lambda=0$, the operator $L=-W$ delivers the decomposition,
\begin{align}
	0=\lambda\geq f_r(-W) \ \Rightarrow \ W_{-W}=0-(-W)=W.
\end{align}
It follows that $W_L$ is a SN-$r$ witness, and any SN-$r$ witness can be written as given in Eq.~(\ref{Eq:SNWitDec}).
This means the set $\mathcal W_r$ and the set which is defined by operators of the form in Eq.~(\ref{Eq:SNWitDec}) are identical.

Moreover, the decomposition in Eq.~(\ref{Eq:SNWitDec})~and~(\ref{Eq:SNPos}) proves us the fact that optimized witnesses are necessary and sufficient for the detection of a SN.
Summing up all these findings, we can formulate optimized, necessary and sufficient conditions in terms of arbitrary operators.
\begin{theorem}\label{Theo:Condition}
A quantum state $\varrho$ has a SN greater than $r$, if and only if there exists $L\in{\sf Herm}(\mathcal H)$ with
\begin{align*}
	\langle L\rangle={\rm Tr}\, \varrho L>f_r(L).
\end{align*}
\end{theorem}
\paragraph*{Proof.}
The first part of the proof has been done above with the general structure of a witness in Eq.(\ref{Eq:SNWitDec}).
We use this general form of an optimized SN-$r$ witness $W_L=f_r(L)\mathbb I\otimes\mathbb I-L$.
In addition, we use the fact that a witness must exist:
\begin{align*}
	0>{\rm Tr}\,\sigma W_L=f_r(L){\rm Tr}\,\varrho-{\rm Tr}\, \varrho L \Leftrightarrow {\rm Tr}\, \varrho L >f_r(L).\ \blacksquare
\end{align*}
This condition means that the expectation value of $\varrho$ exceeds the boundary given by all SN-$r$ states.
The formulation of the theorem could also be given by
\begin{align}
	{\rm Tr}\, \varrho L<\inf \{\langle \psi|L|\psi\rangle: |\psi\rangle\langle\psi|\in\mathcal S_r^{\rm (pure)}\},
\end{align}
if we use the minimal value instead of the maximal value in Definition~\ref{Def:MaxExp}.

In conclusion, we have obtained SN-$r$ conditions in terms of arbitrary operators.
So far, only some special examples of SN-$r$ witnesses have been known, so that a general approach was unknown.
This shortcoming we have resolved in terms of the condition in Theorem~\ref{Theo:Condition}.
Such a method is of some general interest, since the identification of the SN of a quantum state delivers the structure and the quantification of entanglement in the notion of global superpositions.
In the following, we will focus on the determination of $f_r$, which is desired for a general characterization of entanglement through the SN.

\section{The $r$-Schmidt eigenvalue problem}\label{Sec:SNEVAL}
Now, we consider the function $f_r$ as given in Definition~\ref{Def:MaxExp}.
The value of $f_r(L)$ denotes the maximal expectation value of a given operator $L$ under all SN-$r$ states.
First, we consider all extrema -- not only the global maximum.
Obviously the largest extrema is the desired value of $f_r$.
The included optimization problem will lead us to equations which are of an algebraic nature.
This means we transform a optimization problem to an algebraic system of equations.

\subsection{The optimization problem}
First of all, let us consider a weaker decomposition than the Schmidt decomposition given in Eq.~(\ref{Eq:SchmDec}).
A pure quantum state $|\psi\rangle$ has a SN less or equal to $r$, if it can be written as
\begin{align}\label{Ee:SchmDecWeak}
	|\psi\rangle=\sum_{k=1}^r |x_k,y_k\rangle,
\end{align}
with $\langle\psi|\psi\rangle=1$.
Here, the product states $|x_k, y_k\rangle$ are not necessarily orthonormal or linear independent.
Some of them can also be the zero.
Thus, $|\psi\rangle$ might have a Schmidt rank less than $r$.
But in any case it follows that $|\psi\rangle\langle\psi|\in\mathcal S_r^{\rm (pure)}$.

The optimization problem given in Definition~\ref{Def:MaxExp} is the following.
For a given $L\in{\sf Herm}(\mathcal H)$, we want to find the maxima or minima -- in general extrema -- of $\langle \psi|L|\psi\rangle$.
For a quantum state $|\psi\rangle\langle\psi|\in\mathcal S_r^{\rm (pure)}$ the expectation value is given as
\begin{align}
	\nonumber G(|\psi\rangle)=&G(|x_1\rangle,\dots,|x_r\rangle,|y_1\rangle,\dots,|y_r\rangle)\\
	=&\langle \psi|L|\psi\rangle \to G_{\rm opt},
\end{align}
where $G_{\rm opt}$ denotes an optimal value.
We also make use of the normalization condition
\begin{align}
	C(|\psi\rangle)=&\langle \psi|\psi\rangle-1=\langle \psi|\mathbb I\otimes\mathbb I|\psi\rangle-1\equiv 0.\label{Eq:CondNorm}
\end{align}

An optimization problem under a certain condition can be solved with the method of Lagrangian Multipliers $g$.
This means for all $k$ holds
\begin{align}
	\label{Eq:LagrMult1}\frac{\partial G}{\partial \langle x_k|}-g\frac{\partial C}{\partial \langle x_k|}&=0,\\
	\label{Eq:LagrMult2}\frac{\partial G}{\partial \langle y_k|}-g\frac{\partial C}{\partial \langle y_k|}&=0.
\end{align}
Using $|\psi\rangle$ in the form of Eq.~(\ref{Ee:SchmDecWeak}), we obtain from Eqs.~(\ref{Eq:LagrMult1})~and~(\ref{Eq:LagrMult2}) for all $k$
\begin{align}
	\label{Eq:LagrMult3}\sum_j \langle y_k|L|x_j,y_j\rangle=&g\sum_j \langle y_k|y_j\rangle|x_j\rangle,\\
	\label{Eq:LagrMult4}\sum_j \langle x_k|L|x_j,y_j\rangle=&g\sum_j \langle x_k|x_j\rangle|y_j\rangle.
\end{align}
The Lagrangian Multiplier $g$ can be obtained by summing up Eq.~(\ref{Eq:LagrMult3}) -- summation given by $\sum_k \langle x_k|\dots$ -- and using the condition $C$ in Eq.~(\ref{Eq:CondNorm}):
\begin{align}
	\nonumber g\cdot 1&=g \sum_{k,j}\langle x_k,y_k|x_j,y_j\rangle\\
	&=\sum_{k,j}\langle x_k,y_k|L|x_j,y_j\rangle=G_{\rm opt}.
\end{align}
This is already the algebraic conversion of the optimization problem.

Further on, let us simplify and rewrite these equations.
Therefore we may consider the following definitions
\begin{align}
	&|\vec x\rangle=\left(\begin{array}{c}|x_1\rangle \\ \vdots \\ |x_r\rangle \end{array}\right), \quad |\vec y\rangle=\left(\begin{array}{c}|y_1\rangle \\ \vdots \\ |y_r\rangle \end{array}\right),\\
	&L_{\vec x}=\left( {\rm tr}_A L\left[|x_k\rangle\langle x_j|\otimes \mathbb I\right]  \right)_{k,j=1,\dots,r},\\
	&L_{\vec y}=\left( {\rm tr}_B L\left[\mathbb I\otimes|y_k\rangle\langle y_j|\right]  \right)_{k,j=1,\dots,r}.
\end{align}
Analogously to the operators $L_{\vec x}$ and $L_{\vec y}$, we define the {\em pseudo-metrics} $\mathbb I_{\vec x}$ and $\mathbb I_{\vec y}$.
From Eqs.~(\ref{Eq:LagrMult3})~and~(\ref{Eq:LagrMult4}) we obtain the first set of equations for the SN-$r$ optimization problem,
\begin{align}
	\label{Eq:SESNr1} L_{\vec y}|\vec x\rangle=&g \mathbb I_{\vec y}|\vec x\rangle,\\
	\label{Eq:SESNr2} L_{\vec x}|\vec y\rangle=&g \mathbb I_{\vec x}|\vec y\rangle.
\end{align}
This form generalizes the separability eigenvalue equations, as derived in~\cite{SpeVo1}.
In fact, for $r=1$ Eqs.~(\ref{Eq:SESNr1})~and~~(\ref{Eq:SESNr2}) reduce to the separability eigenvalue equations.

\begin{definition}\label{Def:rSEprobl1}
The $r$-Schmidt eigenvalue problem is defined in Form 1 as
\begin{align*}
	L_{\vec y}|\vec x\rangle=g \mathbb I_{\vec y}|\vec x\rangle\ \text{ and } \ L_{\vec x}|\vec y\rangle=g \mathbb I_{\vec x}|\vec y\rangle.
\end{align*}
We use the abbreviation $r$-SE for $r$-Schmidt eigenvalue.
The vector $|\psi\rangle$ is the $r$-Schmidt eigenvector ($r$-SE vector), and the value $g$ is the $r$-SE for $|\psi\rangle$ of $L$.
\end{definition}
The conversion of the optimization problem to its algebraic form is given in the following theorem which summarizes the above calculations.
\begin{theorem}
	The SN-$r$ state $|\psi\rangle$ delivers an extremal expectation value $g=\langle\psi|L|\psi\rangle$, if and only if it solves the $r$-SE problem defined in
	Definition~\ref{Def:rSEprobl1}.
	$\blacksquare$
\end{theorem}

More sophisticated is the derivation of the second form of these equations.
We note that the SN-$r$ state in its weak form of Eq.~(\ref{Ee:SchmDecWeak}) must also have a {\em strict} Schmidt decomposition, see Eq.~(\ref{Eq:SchmDec}).
Obviously the optimization and the optimal value $G_{\rm opt}=g=\langle\psi|L|\psi\rangle$ does not depend on the form of a decomposition of the pure state.
Let us note that the $r$ in Eq.~(\ref{Ee:SchmDecWeak}) is not necessarily the Schmidt rank $r\geq r(\psi)$ of $|\psi\rangle$.
However, due to the independence of the decomposition and $|\psi\rangle\langle\psi|\in\mathcal S_{r(\psi)}^{\rm (pure)}\subset\mathcal S_{r}^{\rm (pure)}$ Hence, we relabel formally
\begin{align}
	|x_k\rangle\to|e_k\rangle, \  |x_k\rangle\to\lambda_k|f_k\rangle \ \text{ and } \ r(\psi)\to r.
\end{align}

Now, let us reconsider which information is included in Eqs.~(\ref{Eq:LagrMult3})~and~(\ref{Eq:LagrMult4}), and rewrite them as
\begin{align}
	\label{Eq:LagrMult10}\langle e_k|L|\psi\rangle=&g \langle e_k|\psi\rangle=g\lambda_k|f_k\rangle,\\
	\label{Eq:LagrMult11}\lambda_k\langle f_k|L|\psi\rangle=&g\lambda_k \langle f_k|\psi\rangle=g\lambda_k^2|e_k\rangle.
\end{align}
We use the notion $\langle e_k|\psi\rangle\in\mathcal H_B$ and $\langle f_k|\psi\rangle\in\mathcal H_A$ for the projection of the state $|\psi\rangle$ on the $|e_k\rangle$ and $|f_k\rangle$ component, respectively.
Note that $\lambda_k>0$.
Thus, for each $k$, we can multiply Eq.~(\ref{Eq:LagrMult11}) with $\lambda_k^{-1}$.
Equations~(\ref{Eq:LagrMult10})~and~(\ref{Eq:LagrMult11}) yield for $k,k'=1,\dots,r$
\begin{align}
	\label{Eq:LagrMult5}\langle e_k,f_{k'}|L|\psi\rangle=g\lambda_k\delta_{k,k'}.
\end{align}

On the other hand, let us consider the action of $L$ on the pure SN-$r$ state $|\psi\rangle$.
This delivers the (not normalized) result $|\psi'\rangle$,
\begin{align}
	L|\psi\rangle=|\psi'\rangle=g|\psi\rangle+|\chi\rangle,\label{Eq:LinOnPsi}
\end{align}
with the decomposition of $|\psi'\rangle$ into a parallel component, $g|\psi\rangle$, and an orthogonal one, $|\chi\rangle$.
Now on the subsystems $A$ and $B$ act $|e_k\rangle$ and $|f_{k'}\rangle$, respectively,
\begin{align}
	\label{Eq:LagrMult7}\langle e_k,f_{k'}|L|\psi\rangle=&\langle e_k, f_{k'}|\psi'\rangle,
\end{align}
We combine Eqs.~(\ref{Eq:LagrMult5})~--~(\ref{Eq:LagrMult7}) and obtain
\begin{align}
	\langle e_k, f_{k'}|\psi'\rangle&=g\langle e_k,f_{k'}|\psi\rangle+\langle e_k,f_{k'}|\chi\rangle=g\langle e_k,f_{k'}|\psi\rangle,
\end{align}
for all $k,k'=1,\dots,r$.
We conclude that $|\chi\rangle$ is orthogonal to each $|e_k,f_{k'}\rangle$ for $k,k'=1,\dots,r$ which leads to
\begin{align}
	\label{Eq:VecChi}|\chi\rangle=\sum_{k=r+1}^{d_A}\sum_{k'=r+1}^{d_B}\chi_{k,k'}|e_k,f_{k'}\rangle.
\end{align}
This means that $|\chi\rangle$ is not only orthogonal to $|\psi\rangle$, but also to all linear combinations of $|e_k,f_{k'}\rangle\in\mathcal H_A\otimes\mathcal H_B$ of the Schmidt decomposition of $|\psi\rangle$ and $k,k'=1,\dots,r$.
We denote this property of Eq.~(\ref{Eq:VecChi}) as {\em bi-orthogonality}.

From these considerations follows another algebraic conversion of the optimization problem as
\begin{align}
	L|\psi\rangle=g|\psi\rangle+|\chi\rangle,
\end{align}
with a bi-orthogonal vector $|\chi\rangle$ as it is given in Eq.~(\ref{Eq:VecChi}).
We define the second algebraic form of the optimization problem for a state $|\psi\rangle$ with maximally $r$ global superpositions of factorizable states.
\begin{definition}\label{Def:rSEprobl}
The $r$-Schmidt eigenvalue problem is defined in Form 2 as:
\begin{align*}
	L|\psi\rangle=g|\psi\rangle+|\chi\rangle,
\end{align*}
with a bi-orthogonal perturbation $|\chi\rangle$.
\end{definition}
Due to the fact that Definitions~\ref{Def:rSEprobl1}~and~\ref{Def:rSEprobl} solve the same optimization problem, Form 1 and Form 2 of the $r$-SE problem are equivalent.
\begin{theorem}
	The SN-$r$ state $|\psi\rangle$ delivers an extremal expectation value $g=\langle\psi|L|\psi\rangle$, if and only if it solves the $r$-SE problem defined in
	Definition~\ref{Def:rSEprobl}.
	$\blacksquare$
\end{theorem}

\subsection{The $r$-SE problem -- Preliminary conclusions}
So far, we have derived two forms of the $r$-SE problem.
Our initial intention was to determine the value of the function $f_r(L)$ for a given linear operator $L$.
The $r$-SE $g$ is the extremal value of the function $G$,
\begin{align}
	G(|\psi\rangle)=\langle\psi|L|\psi\rangle=G_{\rm opt}=g.
\end{align}
Under all extremal values the global maximum can be found as
\begin{corollary}
	$f_r(L)=\max \{g: g\text{ $r$-SE value of $L$}\}$.
	$\blacksquare$
\end{corollary}

In some cases $L$ cannot be used for the identification of SN-$r$ states.
Obviously holds that $L$ is not suitable in such a case, if and only if $f_r(L)=f_{d}(L)$, cf. Eqs.~(\ref{Eq:SNWit})~and~(\ref{Eq:WitWit}) and $d$ denoting the maximal possible Schmidt rank.
This means, no quantum state with an arbitrary SN exceeds the maximal expectation value of $L$ for SN-$r$ states.
Thus, the test operator $L$ for the identification of the SN must have the following property.
The eigenspace of the largest eigenvalue does not include a vector with a Schmidt rank less or equal to $r$.
For $r=1$ this is related to the range criterion for quantum states~\cite{rangeCrit}.

It is clear that the bi-orthogonal perturbation $|\chi\rangle$ has a Schmidt decomposition as well.
Due to the bi-orthogonal property it can be given as
\begin{align}
	|\chi\rangle=\sum_{k=r+1}^{d} \tilde\lambda_k|e_k,f_k\rangle.
\end{align}
Now the $r$-SE equation in Definition~\ref{Def:rSEprobl} can be rewritten as
\begin{align}
	L\sum_{k=1}^r \lambda_r |e_k,f_k\rangle=g\sum_{k=1}^r \lambda_r |e_k,f_k\rangle+\sum_{k=r+1}^{d} \tilde\lambda_k|e_k,f_k\rangle.\label{Eq:Form3}
\end{align}
Thus, we can write a necessary condition for a solution $|\psi\rangle$ of the $r$-SE problem.
\begin{corollary}\label{Cor:CommonSDec}
	A $r$-SE vector $|\psi\rangle$ and the vector $L|\psi\rangle$ have a Schmidt decomposition with the same local unitary operation $U\otimes V$,
	\begin{align*}
		(U\otimes V)|\psi\rangle=&\sum_{k=1}^r \lambda_k |k,k\rangle \\
		\text{ and } \ (U\otimes V)L|\psi\rangle=&\sum_{k=1}^{d} \tilde\lambda_k |k,k\rangle. \ \blacksquare
	\end{align*}
\end{corollary}
This condition is only necessary.
It would also be sufficient, if $\tilde\lambda_k=g\lambda_k$ for $k=1,\dots,r$.
But Corollary~\ref{Cor:CommonSDec} in its given form is already a quite strong restriction for finding $r$-SE solutions.
We will see this later when we apply our method.

\subsection{Relation to the eigenvalue problem}
We consider the $r$-SE problem for all possible $r=1,\dots,d$.
In addition, let us consider the {\em (ordinary)} eigenvalue problem of $L\in{\sf Lin}(\mathcal H,\mathcal H)$ in linear algebra,
\begin{align}
	L|\psi\rangle=g|\psi\rangle,
\end{align}
with the eigenvector $|\psi\rangle$ and the eigenvalue $g$.
It is clear that in this case there is a trivial perturbation $|\chi\rangle=0$.
Thus, eigenvalues and eigenvectors, $g$ and $|\psi\rangle$, are also solutions of the $r$-SE problem for $r=r(\psi)$.
The other way around, we conclude for $r=d$ that no bi-orthogonal perturbation can remain.
Therefore, we obtain in this case that the $d$-SE problem and the eigenvalue problem are identical.

Now, let us consider local invertible operations.
These are operations which can be written as $S\otimes T$, with $S$ and $T$ invertible operations, $S^{-1}S=\mathbb I$ and $T^{-1}T=\mathbb I$.
These operations cannot change the amount of entanglement, given by the SN, for any quantum states~\cite{Leinaas,EntMeasures}.
Now we transform the $r$-SE problem of the operator $L$ given in Definition~\ref{Def:rSEprobl} to the $r$-SE of an operator $L'$
\begin{align}
	L'&=\left(S\otimes T\right) L\left(S\otimes T\right)^{-1},\\
	L'|\psi'\rangle&=g'|\psi'\rangle+|\chi'\rangle,
\end{align}
This can be done by multiplying the $r$-SE problem of $L$ with $S\otimes T$,
\begin{align}
	\nonumber\left(S\otimes T\right)L|\psi\rangle&=\left(S\otimes T\right)L\left(S\otimes T\right)^{-1}\left(S\otimes T\right)|\psi\rangle
	\\&=g\left(S\otimes T\right)|\psi\rangle+\left(S\otimes T\right)|\chi\rangle,\label{Eq:41}
\end{align}
We obtain that the $r$-SE values remain, and the vectors are transformed by $S\otimes T$,
\begin{align}
	g=g', \ |\psi'\rangle=\left(S\otimes T\right)|\psi\rangle,\ \text{ and } \ |\chi'\rangle=\left(S\otimes T\right)|\chi\rangle.
\end{align}
\begin{corollary}\label{Cor:LocTrans}
	Let $L'$ be a locally transformed operator $L'=\left(S\otimes T\right) L\left(S\otimes T\right)^{-1}$,
	$g$ is a $r$-SE value and $|\psi\rangle$ a is $r$-SE vector of $L$.
	It follows that
	\begin{align*}
	g=g' \ \text{ and } \ |\psi'\rangle=\left(S\otimes T\right)|\psi\rangle
	\end{align*}
	are $r$-SE value and $r$-SE vector of $L'$, respectively.
\end{corollary}
\paragraph*{Proof.}
Any local invertible map can be decomposed in terms of unitary and diagonal matrices, e.g.. by the singular value decomposition.
First let us consider the local unitary $U\otimes V$, $(U\otimes V)^{-1}=U^\dagger\otimes V^\dagger$.
We use
\begin{align*}
	U\otimes V|e_k,f_l\rangle=&|u_k,v_l\rangle \quad\text{(orthonormal basis)}\\
	L'=&(U\otimes V)L(U^\dagger\otimes V^\dagger),
\end{align*}
with the $r$-SE problem
\begin{align*}
	L\sum_{k=1}^{r}\lambda_k|e_k,f_k\rangle=&g\sum_{k=1}^{r}\lambda_k|e_k,f_k\rangle+\sum_{k=r+1}^{d}\tilde\lambda_k|e_k,f_k\rangle.
\end{align*}
It follows from Eq.~(\ref{Eq:41}) that
\begin{align*}
	L'\sum_{k=1}^{r}\lambda_k|u_k,v_k\rangle=&g\sum_{k=1}^{r}\lambda_k|u_k,v_k\rangle+\sum_{k=r+1}^{d}\tilde\lambda_k|u_k,v_k\rangle.
\end{align*}
In addition we consider local, diagonal, and invertible maps,
\begin{align*}
	D_A\otimes D_B|e_k,f_l\rangle=&d_{k,A}d_{l,B}|e_k,f_l\rangle \quad\text{($d_{k,A},d_{l,B}\neq0$)}\\
	L'=&(D_A\otimes D_B)L(D_A^{-1}\otimes D_B^{-1}),
\end{align*}
and we conclude for the transformed $r$-SE problem
\begin{align*}
	L'\sum_{k=1}^{r}d_{k,A}d_{k,B}\lambda_k|e_k,f_k\rangle=&g\sum_{k=1}^{r}d_{k,A}d_{k,B}\lambda_k|e_k,f_k\rangle\\&+\sum_{k=r+1}^{d}d_{k,A}d_{k,B}\tilde\lambda_k|e_k,f_k\rangle.
\end{align*}
The Schmidt coefficients change, $|\psi\rangle\mapsto|\psi'\rangle$, but the $r$-SE value remains invariant. $\blacksquare$

For the {\em (ordinary)} eigenvalue problem in linear algebra an invertible -- in general global -- transformation $T_{AB}$ can be found for the diagonalization of the matrix.
In general, a matrix $L$ can be transformed into another one by $L'=T_{AB}LT_{AB}^{-1}$.
This transformation changes the eigenvectors in the same way, and the eigenvalues remain invariant.
For the $r$-SE problem we have a related situation which, in addition, reflects the global superposition property of SN $r$ states.

Let us consider a special case of local invertible operations, cf. proof of Corollary~\ref{Cor:LocTrans}.
The same considerations are obviously true for $S\otimes T=U\otimes V$, with $U$ and $V$ being unitary operations.
Consequently, the $r$-SE problem is invariant under local unitary, which is of importance for the quantification of entanglement, cf. e.g.~\cite{book2,book3}.

\section{Solutions of the $r$-SE problem}\label{Sec:SESol}
So far we have considered general properties of the $r$-SE problem.
We have compared our method with the eigenvalue problem in linear algebra.
Now let us apply our method to some examples.
Starting from low rank matrices, we explain how the $r$-SE problem can also be solved for higher rank operators.
We consider an approach connecting the spectral decomposition with the $r$-SE problem.
Our SN-$r$ conditions will be tested for some examples in continuous variable systems.

\subsection{One-dimensional projections}\label{SS:ODP}
For simplicity let us here and in the following restrict to Hilbert spaces with $\dim\ \mathcal H_A=\dim\ \mathcal H_B=d$.
We consider a one dimensional projection $L=|\phi\rangle\langle\phi|$, with $1\leq r(\phi)\leq d$,
\begin{align}
	|\phi\rangle=\sum_{k=1}^{r(\phi)} \kappa_k|k,k\rangle,
\end{align}
$\kappa_k\in\mathbb C\setminus\{0\}$, and the normalization $\sum_{k} |\kappa_k|^2=1$.
In addition, let us assume that the coefficients $\kappa_k$ are ordered as $|\kappa_1|\geq|\kappa_2|\geq\dots\geq|\kappa_{r(\phi)}|$.
It is also useful to define $\kappa_{r(\phi)+1}=\dots=\kappa_d=0$.

We aim to solve the $r$-SE problem for $L$.
First, there are a lot of trivial solutions given by orthogonal eigenvectors
\begin{align}
	L|\psi\rangle=0|\psi\rangle+|\chi\rangle, \ \text{ for } \
	|\psi\rangle\perp|\phi\rangle \ \text{ and } \ |\chi\rangle=0.
\end{align}
An example is $|\psi\rangle=(|2,1\rangle+|1,2\rangle)/\sqrt 2$ for $r=2$.

The non-trivial solutions, $g\neq0$, are more interesting for obtaining $f_r(L)$.
Applying Corollary~\ref{Cor:CommonSDec} we conclude that we only need to consider states $|\psi\rangle$ with a decomposition as
\begin{align}
	|\psi\rangle=\sum_{k=1}^{d} \lambda_{k} |k,k\rangle,
\end{align}
with $\lambda_k\in\mathbb C$.
The Schmidt rank of $|\psi\rangle$ is the number of coefficients with $\lambda_k\neq0$.
For a Schmidt rank less than $d$ some of the $\lambda_k$ have to be zero.

Thus, we obtain by applying the projection, $L|\psi\rangle=g|\psi\rangle+|\chi\rangle$,
\begin{align}
	\nonumber L|\psi\rangle=&\langle\phi|\psi\rangle|\phi\rangle=\left(\sum_{l=1}^{d} \kappa_l^\ast\lambda_l\right)\sum_{k=1}^{d}\kappa_k|k,k\rangle\\
	=&g\sum_{k=1}^d \lambda_{k} |k,k\rangle + \sum_{k=1}^{d} \tilde\lambda_k|k,k\rangle.\label{Eq:CoeffProj}
\end{align}
A closer look on Eq.~(\ref{Eq:CoeffProj}) delivers all the possible solutions:
\begin{align}\label{Eq:SchmCoeffSEVec}
	\lambda_k=c\kappa_k\text{ or } \lambda_k=0,
\end{align}
with a constant number $c=const.\neq0$.
The bi-orthogonality property between $|\psi\rangle$ and $|\chi\rangle$ delivers that from $\lambda_k\neq0$ follows $\tilde\lambda_k=0$.

For the case $r=2$ we obtain all solutions as
\begin{align}
	|\psi_{k,l}\rangle&=\frac{1}{\sqrt{|\kappa_k|^2+|\kappa_l|^2}}\left(\kappa_k |k,k\rangle+\kappa_l|l,l\rangle\right),\\
	\nonumber g_{k,l}&=\langle\psi_{k,l}|L|\psi_{k,l}\rangle=\frac{(|\kappa_k|^2+|\kappa_l|^2)^2}{|\kappa_k|^2+|\kappa_l|^2}=|\kappa_k|^2+|\kappa_l|^2,
\end{align}
for $k,l=1,\dots,d$ and $k\neq l$.
The maximal $2$-SE vector and value are
\begin{align}
	|\psi_{1,2}\rangle&=\frac{1}{\sqrt{|\kappa_1|^2+|\kappa_2|^2}}\left(\kappa_1 |1,1\rangle+\kappa_2|2,2\rangle\right),\\
	\nonumber g_{1,2}&=\langle\psi|L|\psi\rangle=\frac{(|\kappa_1|^2+|\kappa_2|^2)^2}{|\kappa_1|^2+|\kappa_2|^2}=|\kappa_1|^2+|\kappa_2|^2.
\end{align}

In conclusion, every $|\psi\rangle$ -- with $r$ Schmidt coefficients $\kappa_k=\lambda_k\neq0$ -- and the same Schmidt decomposition as $|\phi\rangle$ solves the $r$-SE problem.
The $r$-SE value is given as $g=|\langle\phi|\psi\rangle|^2$.
Using the Schmidt coefficients with the largest absolute value, we obtain $f_r(L)=\sum_{k=1}^r |\kappa_k|^2$.

For example, let us use $r(\phi)=d=\infty$ and the projection operator $L$ constructed from
\begin{align}
	|\phi\rangle=|q\rangle=\sqrt{1-|q|^2}\sum_{k=0}^\infty q^k |k,k\rangle,
\end{align}
for $0<|q|=\epsilon<1$ (note that the index starts from 0).
This is the two-mode squeezed-vacuum state.
In this case we obtain the maximal solution of the $r$-SE problem as
\begin{align}
	|\psi_r\rangle&=\sqrt{\frac{1-|q|^{2}}{1-|q|^{2r}}}\sum_{k=0}^{r-1} q^k|k,k\rangle,\\
	g_r&=|\langle\psi_r|\phi\rangle|^2
	=1-|q|^{2r}=f_r(|\phi\rangle\langle\phi|).
\end{align}
For the quantum state $\rho_0=|\epsilon\rangle\langle\epsilon|$ ($q=\epsilon$) we can identify the entanglement with an arbitrary SN by
\begin{align}\label{Eq:TMSVCond}
	{\rm Tr}\,\rho_0 L=1>1-|q|^{2r}.
\end{align}
This optimal violation, ${\rm Tr}\,\rho_0 L=f_\infty(L)$, does not depend on the amount of squeezing $|q|$.
Even in the case of a minimal squeezing, $|q|\ll 1$, the condition Eq.~(\ref{Eq:TMSVCond}) proves that the state $\rho_0$ has a SN larger than any $r$.

Another example could be $r(\phi)=d$ and equally distributed Schmidt coefficients,
\begin{align}
	|\phi\rangle=\frac{1}{\sqrt d}\sum_{k=1}^d |k,k\rangle.
\end{align}
In this case we obtain the maximal solution of the $r$-SE problem for $L=|\phi\rangle\langle\phi|$ as
\begin{align}
	|\psi_r\rangle&=\frac{1}{\sqrt r}\sum_{k=1}^{r} |k,k\rangle,\\
	g_r&=|\langle\psi_r|\phi\rangle|^2
	=\frac{r}{d}=f_r(|\phi\rangle\langle\phi|).
\end{align}
In conclusion, we can formulate the following necessary criteria.
A quantum state $\varrho$ has a Schmidt rank greater than $r$ if
\begin{align}
	\langle\phi|\varrho|\phi\rangle>\sum_{k=1}^r |\kappa_k|^2,
\end{align}
with $|\kappa_k|$ ($k=1,\dots,r$) being the $r$ largest Schmidt coefficients of $|\phi\rangle$.

\subsection{Higher rank operators}
An arbitrary Hermitian operator $L$ can be decomposed in terms of one-dimensional projectors.
In the following, we generalize our results for one-dimensional projections to classes of higher rank operators.
It is of advantage to consider all one dimensional projections as a local transformation of a given state with maximal Schmidt rank.
We start from pure vectors $|\psi\rangle$ and $|\Phi\rangle$
\begin{align}
	|\psi\rangle=\sum_{k=1}^{d}\sum_{l=1}^{d} \psi_{k,l}|k,l\rangle,\ \text{ and } \
	|\Phi\rangle=\sum_{n=1}^{d}|n,n\rangle.
\end{align}
The linear map $M_\psi=\sum_{k,l}\psi_{k,l}|k\rangle\langle l|$ yields the desired property
\begin{align}
	|\psi\rangle=\left(M_\psi\otimes\mathbb I\right)|\Phi\rangle=\left(\mathbb I\otimes M_\psi^T\right)|\Phi\rangle.
\end{align}
Note that the singular value decomposition of $M_\psi$ resembles the Schmidt decomposition of $|\psi\rangle$~\cite{book1}.
This rewriting delivers that the scalar product of two states $|\psi\rangle$ and $|\xi\rangle$ is the scalar product of the corresponding matrices
\begin{align}
	\langle\xi|\psi\rangle={\rm Tr}\ M_\xi^\dagger M_\psi.
\end{align}
A general Hermitian operator $L\in{\sf Herm}(\mathcal H)$ can be decomposed as
\begin{align}
	\nonumber L=&\sum _k L_k |\psi_k\rangle\langle\psi_k|
	\\=&\sum_k L_k \left(M_k\otimes\mathbb I\right)|\Phi\rangle\langle\Phi|\left(M_k^\dagger\otimes\mathbb I\right),
\end{align}
with $L_k\in\mathbb R$.
An example would be the spectral decomposition of $L$ with the eigenvalues $L_k$ and the eigenvectors $\left(M_k\otimes\mathbb I\right)|\Phi\rangle$.
We apply $L$ on a given state $(M_\psi\otimes \mathbb I)|\Phi\rangle$ which yields
\begin{align}
	\nonumber L(M_\psi\otimes\mathbb I)|\Phi\rangle=&\sum_k L_k \left({\rm Tr}\left[ M_k^\dagger M_\psi\right]\right) (M_k\otimes\mathbb I)|\Phi\rangle\\
	=&\left(\sum_k L_k {\rm Tr}\left[ M_k^\dagger M_\psi\right] M_k\right)\otimes\mathbb I|\Phi\rangle.
\end{align}
From Corollary~\ref{Cor:CommonSDec} follows that $M_\psi$ and the resulting matrix $\left(\sum_k L_k {\rm Tr}\left[ M_k^\dagger M_\psi\right] M_k\right)$ have the same singular value (Schmidt) decomposition, if $(M_\psi\otimes\mathbb I)|\Phi\rangle$ is a $r$-SE vector.
Note that the Schmidt rank of the state $|\psi\rangle$ is the rank of the corresponding matrix, $r(\psi)={\rm rank}\,M_\psi$.

Let us consider an example.
May the matrices $M_k$ have a decomposition as $M_k=UD_kV^\dagger$, with unitary operators $U$ and $V$ and a diagonal matrix $D_k$ with complex components.
We have seen, that unitary operations basically do not effect the $r$-SE problem, cf. Corollary~\ref{Cor:LocTrans}.
Hence, we choose $U=\mathbb I=V$.
The effect of this example is that the mapping delivers
\begin{align}
	M_\psi\mapsto\left(\sum_k L_k {\rm Tr}\left[ D_k^\dagger M_\psi\right] D_k\right),
\end{align}
which is already given as a diagonal matrix.
In such a case it follows, that the vector $L(M_\psi\otimes\mathbb I)|\Phi\rangle$ is given in Schmidt decomposition and we can treat this case analogously to those given in Section~\ref{SS:ODP} for low rank operators.
We may choose a decomposition $D_k=\sum_j \kappa_{k,j} |j\rangle\langle j|$ (with $\kappa_{k,j}\in\mathbb C$ and $\sum_j |\kappa_{k,j}^2|=1$), and the operator $L$ as
\begin{align}
	\nonumber L=&\sum_{k} L_k D_k\otimes\mathbb I |\Phi\rangle\langle\Phi|D_k^\ast\otimes\mathbb I\\
	\nonumber =&\sum_{k,m,n} L_k \kappa_{k,m}\kappa_{k,n}^\ast |m,m\rangle\langle n,n|\\
	=&\sum_{m,n} \gamma_{m,n} |m,m\rangle\langle n,n|.
\end{align}
The new parameters $\gamma_{m,n}$ are given by
\begin{align}
	\gamma_{m,n}=\sum_k L_k \kappa_{k,m}\kappa_{k,n}^\ast.
\end{align}
It is worth to note that this operator $L$ is symmetric with respect to exchanging the quantum systems.

For obtaining the nontrivial solutions, $g\neq0$, we consider $|\psi\rangle=\sum_n \lambda_n |n,n\rangle$.
This yields
\begin{align}
	L|\psi\rangle=\sum_m \left(\sum_n\gamma_{m,n} \lambda_n\right)|m,m\rangle.
\end{align}
Due to the fact that some $\lambda_k$ can be zero, we obtain the $r$-SE solutions by neglecting some rows and the corresponding columns, cf. Eq.~(\ref{Eq:SchmCoeffSEVec}).
These rows are given by the Schmidt coefficients of $|\psi\rangle$ which are zero, $\lambda_{q}=0$.
This means we restrict to the rows and columns with the index $q_1,\dots,q_r$, and solve the (ordinary) eigenvalue problem of the resulting operator,
\begin{align}
	L(q_1,\dots,q_r)=\sum_{i,j=1}^r \gamma_{q_i,q_j} |q_i,q_i\rangle\langle q_j,q_j|.
\end{align}

The simplest case is $r=1$.
We have to stroke out $d-1$ rows and the corresponding columns.
It immediately yields
\begin{align}
	L(q_1)=\gamma_{q_1,q_1}|q_1,q_1\rangle\langle q_1,q_2|.
\end{align}
The maximal expectation value for all possible choices of $q_1=n=1,\dots,d$ delivers the function $f_1(L)$ as
\begin{align}
	f_1(L)=\max_{n} \gamma_{n,n}.\label{Eq:SN1f1}
\end{align}

In the case $r=2$, we have to consider all principal $2\times 2$ sub-matrices of $\gamma_{m,n}$ with $q_1,q_2=m,n$.
We calculate the maximal eigenvalue, and obtain the function $f_2(L)$ as
\begin{align}
	f_2(L)=\max_{m,n, m\neq n} &\left( \frac{\gamma_{m,m}+\gamma_{n,n}}{2} \right.\label{Eq:SN2f2}
	\\ \nonumber &\,\left.+\frac{\sqrt{(\gamma_{m,m}-\gamma_{n,n})^2+4|\gamma_{m,n}|^2}}{2}\right),
\end{align}
denoting the maximal eigenvalue of all principal $2\times 2$ sub-matrices.
Analogously, the method can be applied to SN-$r$ states by finding the maximal eigenvalue of all $r\times r$ principal sub-matrices of $\gamma_{m,n}$.

\section{Phase randomized two-mode squeezed-vacuum}\label{Sec:Ex}
Let us come back to our initially considered state $\rho_{\delta\varphi}$, cf. Eq.~(\ref{Eq:RhoVarPhi}).
This state was generated by a local phase randomization of a two-mode squeezed-vacuum state, cf. Fig.~\ref{Fig:Exp}.
For witnessing the SN, we already considered the example of a projection defined by a pure two-mode squeezed-vacuum state in Sec.~\ref{SS:ODP}.
This operator was suitable for the detection that the state $\rho_0$, $\delta\varphi=0$, has an infinite SN.
However, we already observed that a total randomization, $\delta\varphi=\pi$, delivers a separable state $\rho_\pi$, cf. Eq~(\ref{Eq:TMSVstate}).

Now, we apply our considerations to the general phase randomized (mixed) state $\rho_{\delta\varphi}$ to answer our initial questions.
Namely, for which values of $\delta\varphi$ the entanglement survives (SN $r>1$), and which values of $\delta\varphi$ guarantee that the state contains an entangled qutrit (SN $r>2$)?
A phase randomization in a single mode state has been experimentally realized, and the non-classicality of such a state has been verified~\cite{Kiesel}.
Here, we consider a similar situation, for two modes, where the desired aspect of non-classicality is the number of global superpositions between these modes.
The method presented in the following also applies to a non-equally distributed phase randomization in both subsystems $A$ and $B$, and an additional amplitude randomization of $\epsilon$, as it occurs in the turbulent atmosphere~\cite{Semenov}.

Here, we chose a test operator $L$ of formally  the same form as the state $\rho_{\delta\varphi}$ in Eq.~(\ref{Eq:RhoVarPhi}),
\begin{align}
	\nonumber L=&\frac{1-|q|^2}{2\delta\varphi}\int_{-\delta\varphi}^{\delta\varphi} d\varphi\, \sum_{m=0}^\infty q^m|m,m\rangle\sum_{n=0}^\infty {q^\ast}^n\langle n,n|\\
	=&\sum_{m,n=0}^\infty \gamma_{m,n} |m,m\rangle\langle n,n|,\\
	\gamma_{m,n}=&(1-\epsilon^2)\epsilon^{m+n}\int_{-\delta\varphi}^{+\delta\varphi} d\varphi\, e^{i\varphi(m-n)},
\end{align}
with $q=\epsilon e^{i\varphi}$ and $0<\epsilon<1$.
For simplicity, we have assumed an equally distributed randomization for the angle, $-\pi<-\delta\varphi\leq\varphi\leq\delta\varphi<\pi$.
Therefore, the corresponding parameters $\gamma_{m,n}$ read as
\begin{align}
	\nonumber\gamma_{m,n}=&\epsilon^{m+n}(1-\epsilon^2)\frac{\sin\left(\delta\varphi[m-n]\right)}{\delta\varphi[m-n]}.\\
	=&(1-\epsilon^2)\epsilon^{m+n}{\rm sinc}\left(\delta\varphi[m-n]\right).
\end{align}

First, we may consider a fully randomized and equally distributed phase for the test operator, $\delta\varphi=\pi$,
\begin{align}
	L=\sum_{m=0}^\infty (1-\epsilon^2)\epsilon^{2m} |m,m\rangle\langle m,m|.
\end{align}
The spectral decomposition of $L$ is given in factorizable states.
Thus, $f_1(L)=f_\infty(L)$, and the phase insensitive operator cannot detect any entanglement.
We already observed this result for the corresponding state $\rho_\pi$ which is separable.

Now we consider a given partially phase randomized operator, $L$.
We obtain from Eq.~(\ref{Eq:SN1f1}) for separable states (SN $r=1$) a maximal expectation value of
\begin{align}
	\nonumber f_1(L)=&\max_{m=0,\dots,\infty} \gamma_{m,m}\\
	=&\max_{m} \epsilon^{2m}(1-\epsilon^2)=(1-\epsilon^2).
\end{align}
The maximal expectation value for two-qubit states (SN $r=2$), cf. Eq.~(\ref{Eq:SN2f2}), yields
\begin{align}
	f_2(L)=&\max_{m,n\in\mathbb N, m\neq n}\left.\frac{1-\epsilon^2}{2}\right(\epsilon^{2m}+\epsilon^{2n}\\
	\nonumber &+\left. \sqrt{
	[\epsilon^{2m}-\epsilon^{2n}]^2+4\epsilon^{2m+2n}{\rm sinc}^2\left(\delta\varphi[m-n]\right)
	}\right)
\intertext{using $n=m+k$ and $k\geq1$ we obtain}
	\nonumber f_2(L)=&\left.\frac{1-\epsilon^2}{2}\max_{m\in\mathbb N}\epsilon^{2m}\max_{k\geq1}\right(1+\epsilon^{2k}\\
	&\left.+\sqrt{[1-\epsilon^{2k}]^2+4\epsilon^{2k}{\rm sinc}^2\left(\delta\varphi[m-n]\right)}\right).
\end{align}
The maximum over $m$ yields $m=0$.
The maximum over $k$ has to be calculated numerically.
Note that $f_1(L)<f_2(L)$, and therefore $L$ can detect entanglement.

Now let us consider the state $\rho_{\delta\varphi}$.
The expectation value $\langle L\rangle_{\delta\varphi}={\rm Tr}\, \rho_{\delta\varphi} L$ of the observable $L$ for the state $\rho_{\delta\varphi}$ reads as
\begin{align}
	\nonumber \langle L\rangle_{\delta\varphi}=&(1-\epsilon^2)^2\sum_{m,n=0}^\infty \epsilon^{2m+2n}{\rm sinc}^2(\delta\varphi[m-n])\\
	\nonumber =&(1-\epsilon^2)^2\sum_{m=0}^\infty\epsilon^{4m}\left(1+2\sum_{k=1}^\infty \epsilon^{2k}{\rm sinc}^2(\delta\varphi k)\right)\\
	=&\frac{1-\epsilon^2}{1+\epsilon^2}\left(1+2\sum_{k=1}^\infty \epsilon^{2k}{\rm sinc}^2(\delta\varphi k)\right).
\end{align}
Again, this can be calculated numerically.
We consider the case $\epsilon=1/3$, which corresponds to a moderate squeezing of about $3\, {\rm dB}$~\cite{Relation}.
The entanglement condition for the choice of an operator $L=\rho_{\delta\varphi}$ can be formulated as
\begin{align}
	{\rm Tr}\,\rho_{\delta\varphi} L-f_r(L)=\langle L\rangle_{\delta\varphi}-f_r(L)>0.
\end{align}
This means a positive number is a verification of a SN greater than $r$.
\begin{center}
\begin{figure}[ht]
\includegraphics*[width=8.5cm]{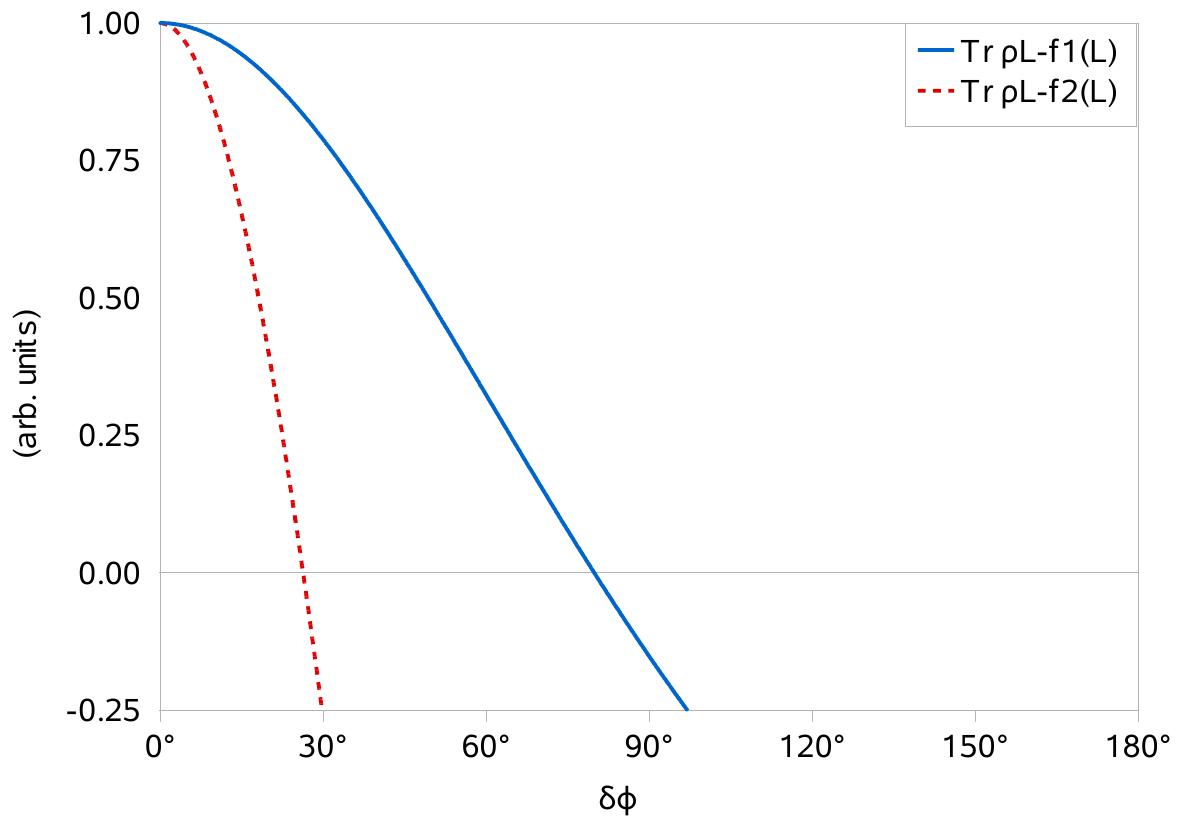}
\caption{
(color online) Here, the detected entanglement for a two-mode squeezed-vacuum state is given.
We choose a squeezing of $3\, {\rm dB}$.
In addition the state is manipulated by an equally distributed phase randomization with the interval of the length $2\delta\varphi$.
The functions are normalized to the maximal value.
The entanglement can be verified until $\delta\varphi=79^\circ$ (blue solid function).
We also identified a greater Schmidt rank than 2 for phase diffusion of $\delta\varphi=25^{\circ}$ (red dashed function).
}\label{Fig:2MSVac}
\end{figure}
\end{center}
In Fig~\ref{Fig:2MSVac}, the identification of entanglement for angles of an equal distribution in phase for $[-\delta\varphi,\delta\varphi]$ is given.
The values are obtained by numerical calculation for the matrix components $\gamma_{m,n}$ with $m,n\leq 100$.
This means that the given positivity, i.e. the entanglement, can only increase for indices $>100$.
Let us stress that we choose an arbitrary operator for the detection of entanglement and a SN greater than 2.
There may exists operators which deliver entanglement also for randomization above the limitations of the chosen operator $L$.

\begin{center}
\begin{figure}[ht]
\includegraphics*[width=8.5cm]{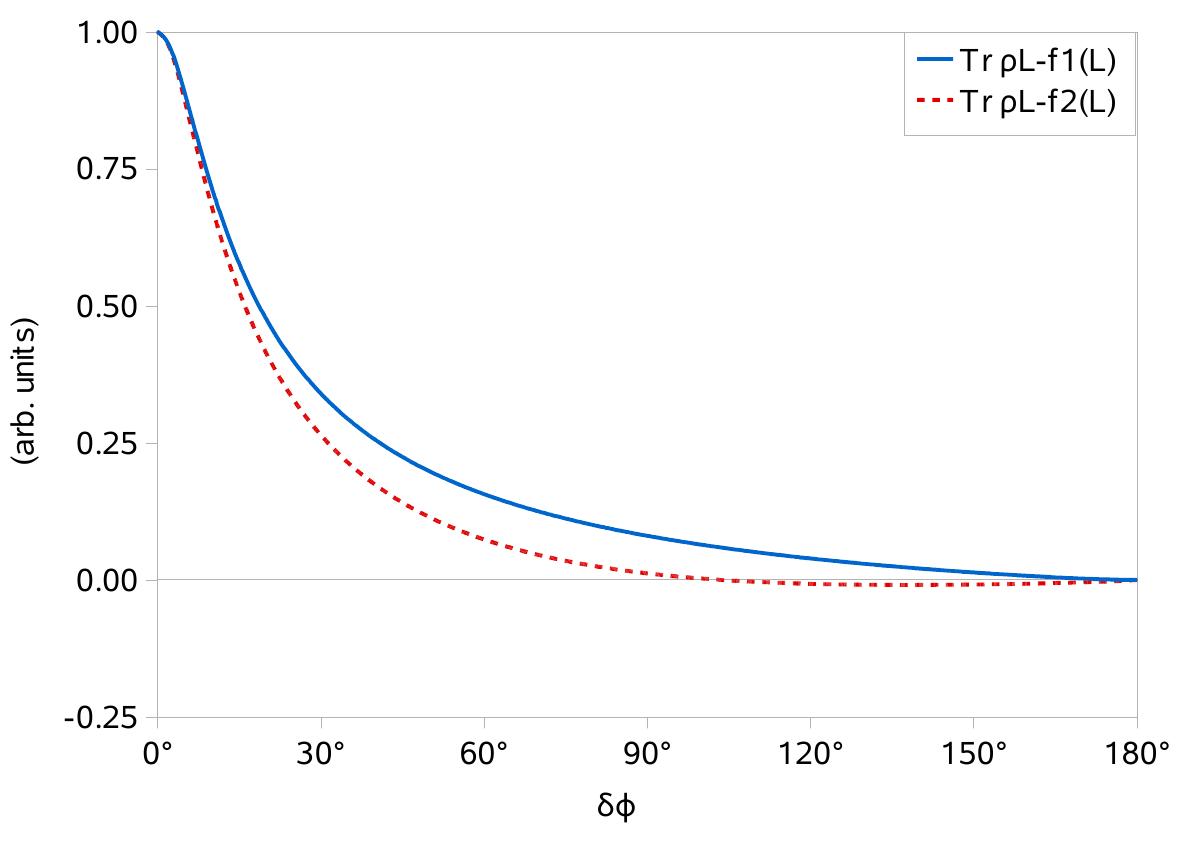}
\caption{
(color online) Here, the detected entanglement is given for higher squeezing, $10\, {\rm dB}$, and the test operator in Eq.~(\ref{Eq:82}).
In the numerical calculation, the entanglement can be verified for a phase randomization up to $\delta\varphi=178^\circ$ (blue solid function).
We also identified a greater SN than 2 for phase diffusion up to $\delta\varphi=102^{\circ}$ (red dashed function).
}\label{Fig:2MSVac2}
\end{figure}
\end{center}

In addition, it is worth to note that a higher squeezing of the state delivers a higher sensitivity.
For a realistic $10\,{\rm dB}$ ($\epsilon=0.82$) squeezing of the input states, cf.~\cite{Schnabel}, we can identify entanglement up to a phase diffusion of $178^\circ$ and a SN greater than 2 close to $102^\circ$, cf. Fig.~\ref{Fig:2MSVac2}.
Here, the chosen test operator $L$ is given as
\begin{align}\label{Eq:82}
	L=\sum_{m,n=0}^{100} {\rm sinc}\left(\delta\varphi[m-n]\right)|m,m\rangle\langle n,n|.
\end{align}
This means that the state $\rho_{\delta\varphi}$ is entangled even for a phase randomization up to to the total phase randomization.
The other interesting point is that the state surely contains a qutrit (SN greater than 2) for a randomization of more then $\pm90^\circ$ in the channel of the receiver.
We note that for such a phase randomization, the squeezing of the single-mode squeezed-vacuum state would vanish.

\section{Summary and Conclusions}\label{Sec:SC}
We have derived optimized, necessary and sufficient conditions for the detection of the Schmidt number of an arbitrary bipartite quantum state.
These conditions have been formulated in terms of arbitrary Hermitian test operators as measurable conditions.
We have shown that the optimization problem leads to the  $r$-Schmidt-eigenvalue equations.
We discussed the properties of the solutions and consequences of these equations in connection with entanglement and its quantification via the Schmidt number.
For example, we have shown which operators can be used for the identification of the Schmidt number, and we have considered the relation to the eigenvalue problem.
We have solved these equations for a wide class of operators, namely for all one dimensional projections and classes of higher rank operators in finite and infinite dimensional systems.
We have shown a direct identification of the Schmidt number for all pure states with our condition.

We also applied our method to the identification of the Schmidt number of mixed quantum states, including those with an infinite rank and infinite Schmidt rank in continuous variable systems.
Examples of test operators are considered, which identify the Schmidt number for a broad class of mixed states.
To illustrate the method, we have studied the example of a phase diffused two-mode squeezed-vacuum state for a squeezing of $3\,{\rm dB}$ and $10\,{\rm dB}$.
We have identified entanglement (Schmidt number greater than one), and we could identify entangled pairs of qutrits in the state (Schmidt number greater than two) for different values of the phase randomization.
Moreover, the influence of the amount of squeezing and the randomization has been considered.

\section*{Acknowledgment}
We are grateful to T. Kiesel for useful discussions.
We are also grateful to the anonymous referee for helpful comments.
This work was supported by the Deutsche Forschungsgemeinschaft through SFB 652.

\end{document}